# Three-dimensional cell culture model for hepatocytes opens a new avenue of real world research on liver

Ting Yao[1], Yi Zhang[1], Mengjiao Lv[1], Guoqing Zang[1], Soon Seng Ng[2]*, Xiaohua Chen[1]*

* Correspondence: soon_seng.ng@kcl.ac.uk, chenxiaohua2000@163.com

[1] Department of Infectious Diseases, Shanghai Jiao Tong University Affiliated Sixth People's Hospital, 600 Yishan Road, Shanghai 200233, China

[2] Centre for Stem Cells and Regenerative Medicine, King's College London, 28th Floor, Tower Wing, Great Maze Pond, London SE1 9AD, United Kingdom

**Electronic word count:** 4984

**Number of figures and tables:** 3 figures; 1 table

**Conflict of interest statement:** No conflict

**Financial support statement:** This work was supported by grants from the National Natural Science Foundation of China (No. 81770589) and Natural Science Foundation of Shanghai (No. 17ZR1421500).

**Authors contributions:** write the paper

**Abstract**

3-demensional (3D) culture model is a valuable in vitro tool to study liver biology, metabolism, organogenesis, tissue morphology, drug discovery and cell-based assays. Compelling evidence suggests that cells cultured in 3D model exhibit superior liver-specific functions over the conventional 2-dimentional (2D) culture in evaluating hepatobiliary drug disposition and drug-induced hepatotoxicity due to the in vivo-like physiological condition recapitulated by 3D model technologies. We will review the attributes of 3D culture model in acquiring relevant liver phenotypes and functionalities, discuss the critical niche factors found to modulate hepatocytes and highlight recent advancements on 3D cell culture technologies to achieve next-level in vitro tool for preclinical study. (Electronic word count of the abstract: 105)

**1. Introduction**

The 3D cell culture is defined as an artificial environment allowing cells to grow and interact in all three spatial dimensions, mimicking the in vivo architecture of liver that comprises of extracellular matrix and nonparenchymal cells [1]. Due to the historical inertia in biotechnology, 2-dimensional (2D) cell culture that relying on cells attaching to surface forming mono layer of confluent cell sheet has instead been widely applied and in fact the predominant liver culture models in preclinical liver studies. However, 2D cell culture model has failed to provide necessary niche factors thus far in sustaining critical cellular phenotype due to the lack of proper intercellular connection in 3D thus generating suboptimal pre-clinical results. For instance, many liver-specific gene expressions and functions of primary human hepatocytes, such as albumin secretion, viral infectivity, and cytochrome

P450 (CYP) enzyme activity [2, 3] are found to be rapidly deteriorating in 2D cell culture upon seeding. These limitations spurred the scientists to explore better culture models. As early as 1912, the 3D culture was first described [4] and opened up an avenue to address the limitation presented in 2D models. In association with the development of biomaterials and biotechnology, many 3D systems with wide range of attributes and features have been proposed for regenerative medicine (showed in **Table 1).**

In healthy condition, rat liver that has undergone surgical removal of up to 70% of liver mass – referred to as partial hepatectomy – restores tissue mass and functions in just a week driven by compensatory proliferation and hyperplasia. While In human, live donor recovery data show that liver mass recovery takes about one year [5]. Unfortunately, this incredible endogenous regenerative capability of liver could be threatened by a series of constant liver assaults including fibrosis and cirrhosis caused by hepatitis infection, drug over dosing, ARLD, NASH, NAFLD, etc [6]. The loss of liver functions could poise detrimental effects on patients because liver plays an important role in a variety of critical functions, including the detoxification of the systemic and portal blood, secretion multiple proteins and bile components [7]. Liver is the main organ in determining the pharmacokinetics of any oral administrated drugs. Most of the drugs that have been withdrawn from the clinical trials along the discovery pipeline or even from the market are partially caused by the use of 2D culture in assessing the dose-dependent efficacy and predict the cytotoxicity of the drug in vitro. For instance, acetaminophen (APAP), the most common household drug for pain killing, is the leading cause for acute liver injury in the West due to overdosing [8]. Some researchers used transcriptome profiling to reveal the mechanisms underlying the high-dose APAP induced mitochondrial dysfunction in the 2D cultured HepG2 model [9]. The lack of expression of this enzyme could explain the insensitivity to APAP in the 2D cultures [10].The failure in drug development is also partially attributed to poor correlation in the use animal models such as rodent, porcine or non-human primate due to the cross-species physiological variation [11-13]. Hepatocellular carcinoma (HCC) is the third leading cause of cancer-related deaths [14, 15] and some researchers believe that high mortality is related to the current limitation of in vitro studies which fail to mimic the human disease microenvironment [16].

This article will first introduce the attributes and niche factors associated with the 3D liver culture system, then focus on some relevant advanced 3D systems and their respective applications.

## 2. Attributes of 3D liver cell culture

The principal goal of developing novel 3D liver culture model is to acquire relevant liver phenotypes and functionalities in vitro. To achieve that, attributes such as microenvironmental architecture, phenotypical sustainability and authentic functionality of the liver must be evaluated in the development of new 3D culture model.

**2.1 Microenvironmental architecture**：Various 3D models have been proposed to recapitulate the microenvironments of liver tissue in vitro in order to achieve authentic cell-cell interactions and cell polarity that are essential for hepatic phenotypes [17-20]. The nature of the interaction with its microenvironment dictates overall shape and cytoarchitecture of hepatocyte which, in turn, are related to the expression of transcription factors and gene programs. For instance, the most widely known model is called sandwich model where a monolayer of hepatocytes is sandwich top-to-bottom by two layers of ECM protein such as collagen I or Matrigel [21]. This model provides minimal cellular interaction needed for hepatocytes to establish hepatic polarity hence enabling bile salt transportation [22]. 3D hydrogel cell encapsulated model is another model heavily employed to facilitate extensive cell-cell and cell-matrix interaction in 3D [23]. Unlike the sandwich model, cell encapsulation model allows higher degree of mobility for cells to navigate. However, the downside is the cells are generally suffering from necrosis due to the suboptimal mass transportation properties in such 3D hydrogel [24] **(Figure 1)**. To circumvent such diffusivity limitation, a series of macroporous 3D scaffolding models such as salt-leaching scaffold, inverse colloidal crystal scaffold and nanofiber scaffolds have been developed to deliver highly porous scaffolds that allow cells to move freely in the 3D microenvironment without compromising their physical integrity.

**2.2 Phenotypical sustainability**: Albeit the incredible regenerative capability in body, hepatocytes harvested from liver tissue dedifferentiate rapidly when culture on 2D model, thus limiting their preclinical potential [25]. Co-culturing stromal cells with hepatocytes in a 3D model that enables secreted matrix to be deposited and organized was found to prolong the hepatic functions over 20 days in culture [26]. On a step forward, a nondegradable PEGdA/RGD hydrogel system was further functionally enhanced by incorporation of hepatocyte aggregates (pucks) as well as supportive stromal cells, resulting in prolonged preservation of hepatic function for 50 days [27]. The longest time for culture systems preserving advanced hepatic functions with primary human hepatocytes was at least 5 months [28].

**2.3 Authentic functionality**: The liver is the main organ for processing drug. The authentic representation of drug metabolism is paramount in preclinical study in evaluating the efficacy and toxicity of the drug. Furthermore, the most potent, or toxic, form of a compound may not be the primary compound but rather one of its metabolites [29]. Due to species-to-species variation in metabolizing drug, a compound's primary metabolite tested in animal study may vary drastically and deem unhelpful in predicting hepatotoxicity in human [30]. Unfortunately, together with the aforementioned rapid deterioration of hepatic functions in vitro, such as the expression of critical cytochrome P450 (CYP) drug metabolizing enzymes, the current mandatory preclinical studies are insufficient to provide an accurate depiction on drug efficacy and safety. The hepatoma cell lines have been proposed and widely used in the preclinical studies due to their superior expension capability and relatively sustainable phenotypes. However, hepatoma cell lines are also known to demonstrate signification variation in their performance and suboptimal functionality compared to PHHs, particularly in detoxification, nitrogen, and carbohydrate metabolism [31]. One study showed ammonia elimination is 64% the rate of elimination of PPH-BALs [32]. Multiple studies have demonstrated the potential of 3D culture model in expressing the authentic metabolic activities compared to the conventional the 2D model [26, 33].

## 3. Vital factors in the 3D cell culture system

### 3.1 Cellular interaction

Hepatocytes in liver organ rely on intricate interaction with its surroundings. Disruption on specific parenchymal cell arrangement, interactions with nonparenchymal cells (NPC) and extracellular matrix is known be the main cause of liver function deterioration observed in 2D model [34, 35].

**3.1.1 Homotypic cell interaction:** Extensive cell–cell interactions is known to be influential to the gene expression profiles and hence functionality [21, 36]. Studies reveal that hepatocytes cultured in spheroids are more powerful than the 2D model in capturing the liver phenotypes and functions in vitro due to extensive cell-cell interaction [10, 21, 24]. In addition, the diffusivity gradient of oxygen and metabolic activity demonstrated by the hepatocytes position from the peripheral to the core of spheroids, when modulated accurately, could be used to model the metabolic zonation in vivo thus making it a more relevant platform to study metabolic response in vitro. On the other hand, the diameter of the spheroid is the most important variable to modulate in order to achieve consistent readout. The configuration is direct contrast to spheroid/cell condensation culture systems where cells at the periphery are more viable and proliferative than cells at the core due to hypoxia and DNA damage [10, 37]. Hussein's research showed the average diameter of the spheroids <180±12μm was sufficient for oxygen diffusion as spheroids above 200μm were reported to become hypoxic in the center [38].

**3.1.2 Extracellular matrix construction:** ECM, that comprises of a mosaic of lipids, proteins and carbohydrates in a complex, heterogeneous and dynamic environment, plays an important role in maintaining the differentiated phenotype of hepatocytes and NPCs [21, 39]. Collagen is the major component of the ECM and it has been widely utilized to culture cells because of its excellent characteristics, including biocompatibility, mechanical strength, degradability and limited immunogenicity [40] and the most common used in 3D model. Matrigel consisting of natural biopolymers: laminin, collagen IV and entactin, as well as various growth factors, has already commercialized [34, 41]. Some researchers indicated that limited matrix formation enhanced hepatic functionality of hepatocytes cultured in vitro [26] and the hepatic differentiation favored a relatively low-density ECM compared with a densely packed ECM surface [42, 43].

**3.1.3 Nonparenchymal cell:** Even though NPCs account for only 20% of the liver mass, they play critical roles in the construction and maintenance of extracellular matrix, as well as mediating cellular function, including transport and metabolism [44]. Major liver NPCs include bile duct epithelial cells, liver sinusoidal endothelial cells (LSEC), hepatic stellate cells (HSC) and Kupffer cells (KC).

LSECs play important roles in maintaining overall hepatic homeostasis and clearance, bioactivation of drugs and other xenobiotics, and they are the target for some types of chemical-induced hepatotoxicities [44]. The LSEC-specific phase 1 enzymes have been less well characterized compared to their epithelial counterparts, but it is clear that they contribute to the metabolism, clearance and bioactivation of endogenous and exogenous substrates [18]. It was revealed that co-culture of primary hepatocyte with LSEC resulted to increase in hepatocyte proliferation [45].

Under normal physiological conditions, HSC are morphologically characterized by their extensive dendrite-like extensions, essentially "embracing" the endothelial cells [46]. This close contact between HSC and their neighboring cells facilitates intercellular communication by the means of soluble mediators and cytokines. HSC can be identified by desmin expression, a typical intermediate filament protein within contractile cells. Mature HSC produce both network and fibrillar collagens (large amounts of type I collagen and lower levels of type III, IV and V collagen), large amounts of elastin and both heparan sulfate proteoglycans (HS-PG) and chondroitin. HSC also produce important cytokines and growth factors for intercellular communication in normal and injured liver [46].

Susceptibility to drugs is different in the healthy liver and in active inflammatory states [47]. Kupffer cells (KC), a type of stellate macrophage reside in the liver, are often targeted as the candidate to study the drug metabolic reaction under inflammation. They play a vital role in immune surveillance of the host and are involved in modulating systemic responses to severe infections and controlling concomitant immune responses via antigen presentation and suppression of the activation and proliferation of T cells [48]. The PHH and KC co-culture models have been developed recently to enable the evaluation of hepatocyte reactions in a pro-inflammatory environment [49, 50]. Notably, this co-culture has demonstrated that inflammation is one of the factors that may increase the sensitivity of hepatic cells to acetaminophen (APAP) induced toxicity [47].

KC is thought the immune cells in generally, but they also play a role in regeneration. In mouse models of acute liver injury, KCs sense injury and become activated, leading to the release of cytokines and chemokines, then the number of macrophages in the liver greatly expands in response to tissue injury, mainly caused by attracting monocytes from the circulation. Although inflammatory may aggravate injury but hepatic macrophages can become the main source of cytokines with anti-inflammatory functions, such as IL-10, IL-4, and IL-13 and cooperate in tissue repair [51]. In mouse models of chronic liver injury, macrophage engulfment of hepatocyte debris induces expression of Wnt3a, which promotes the differentiation of hepatic progenitor cells towards functional hepatocytes favoring parenchymal regeneration [52].

### 3.2 Scaffold construction

Liver is made of a repeated fundamental unit called lobule, in polygonal shape in hexagonal arrangement. Portal triads consisting of the hepatic artery, bile duct, and portal vein are at the corners of the lobule, while the central vein is in the central of the lobule. Plates of parenchymal cells or hepatocytes radiate from the central vein to the perimeter of the lobule, which serves as a microcosm of the major hepatic microenvironments, containing the essential cellular and physiological features that define the unique architecture of the liver tissue. Hepatic plates or cords are generally one hepatocyte thick and are separated from one another by the hepatic sinusoids which are lined by Liver sinusoidal endothelial cells [53, 54]. LSEC plasma membrane is characterized by small pores, or fenestrations, 50–200 nm in diameter that can allow free diffusion of many substances [55]. HSC, reside in the space of Disse-the perisinusoidal space between the basolateral surface of hepatocytes and the anti-luminal side of sinusoidal endothelial cells [56]. KCs are localized within the sinusoidal microvasculature on the luminal side of endothelial cells, however, they have long cytoplasmic extensions that facilitate direct cell-to-cell contact with hepatocytes [57]**(Figure2)**.

With the development of various biotechnology and biomaterials, the constructions of new 3D models come out constantly. The material chosen firstly must be biocompatible with cell growth, and have great performance in chemical and physic stability such as radiation resistance, high reproducibility, less batch-to-batch differences, and good mass transfer capability [58, 59]. Chitosan-gelatin (CG) porous structures, inverted colloidal crystal (ICC) structure, PEGdA/HA, 3D bio-printing all use good biomaterials (some models show in **Figure 3**), and use some methods to produce a porous structure with an excellent surface area to volume ratio facilitating infiltration and communication.

**3.2.1 Decellulerized matrix scaffold:** Natural, non-artificial materials are the closest to the environment in which cells grow. De Kock created decellularized rat livers to evaluate anticancer drugs efficacy [60], which may be the first research exploring the possibility of using a rat liver scaffold to generate 3D model to evaluate anticancer drugs efficacy. Kamal H followed the idea of the De Kock and succeeded in generating decellularized whole rat liver scaffolds that maintained vascular structure and ECM integrity via efficiently removing the cellular and nuclear materials. Next, perfusion of the scaffold with target cells is via the portal vein and the cells were distributed within the scaffold. Successfully, the cells retained their proliferation ability and were functional.

Decellularized rat livers, derived from natural organ, are the ideal structural component of the cell microenvironment which is composed of a sophisticated assembly of collagens, proteoglycans, laminins, elastin, and growth factors. These components represent the basic substances needed for cell growth attachment, growth, and proliferation [38]. The material of natural origin is the closest to the creature, but availability of donor is the challenge.

**3.2.2 Channel-liked scaffolds:** Natural biomaterials such as decellularized liver as scaffold possess perfect innate advantage in providing the right ligands and mircro-environmental cues for cells. However, in additional to limiting sourcing, they known to vary batch to batch hence reproducibility is always an issue for consistent biological readouts and manufacturing. In addition, natural biomaterials are lacking the tunability that is critical for us to manipulate the physical, chemical and biological aspects for highly specialized application. Synthetic biomaterials have thus been proposed to overcome such disadvantages. Polyvinylidene fluoride (PVDF) hollow fibers with massive micropores were used as scaffolds, providing sufficient space and protection for cell growth, meeting the demand of exchanging materials between the inner cells and the outside media. Mixed with collagen solutions, polyvinylidene fluoride (PVDF) hollow fibers will have a rough inner surface to favor adherence, growth, proliferation and functional expressions of the cells. They have two major functions: to gather cells and improve the long-term bioactivity and integrity of the cells. Then cylindroids were formed within the hollow fibers with better adhesion, after cell suspensions and being injected into the hollow fibers [58]. One study showed among the hepatocyte organoids, cylindroids showed better performances over liver-specific functions than spheroids [61].

**3.2.3 Macroporous scaffolds:** Chitosan-gelatin (CG) porous structures were formed by freeze-drying to develop 3D tissues [62]. CG porous structures, one could blend heparan sulfate to chitosan and gelatin and form porous structures to mimic the liver architecture. CG scaffolds have a suitable pore size for infiltration of cells, suitable liver stiffness in hydrated conditions, conducive for adhesion of various cell types [63], good degradation characteristics and scaffolds support matrix synthesis by fibroblasts while minimizing proliferation Morphological changes in HepaRG cells and their migration were observed in 2D cultures over an 8 day culture timeline while in the 3D system were not[64].

Recently, the more advanced system about 3D cell culture is inverted colloidal crystal (ICC) structure. It is also using artificial materials to build the model, but taking into account the anatomy of the liver-3D hexagonally arrayed liver lobules, the functional units of human liver that collectively make up the human liver organ. The internal geometry of the ICC then enables spontaneous formation of heterotypic liver spheroid-like formation within each cavity and interconnected cell growth across adjacent cavities in 3D hexagonally. Some researches indicated that primary human liver fetal cell mixtures were seeded in it and preserved advanced hepatic functions during extended culture time (at least 5 months) [18, 19]. Importantly, this engineered human liver tissue provided proof-of-concept determination of human-specific drug metabolism, remained the ability to support infection with human hepatitis virus for subsequent antiviral drug testing, and facilitated detection of human-specific drug hepatotoxicity associated with late-onset liver failure [65]. ICC scaffolds is with higher degrees of freedom than 2D cultures and other 3D cultures (e.g., bulk hydrogels) [20]. Even if there is no co-cultured cell, free from Matrigel, MSCs, and human umbilical vein endothelial cells (HUVECs), just using a 3D synthetic hydrogel scaffold made by compliant material (PEG) , the ICC is better than 2D, whether in cell phenotype or functional restore [37]. The experimental results show a good application prospect, but as the culture time prolongs, the function and phenotype of the liver will decrease and did not provide hepatocytes with defined stiffness of the 3D parenchymal liver microenvironment.

**3.2.4 Microporous scaffolds:** In vivo, the microenvironment is dynamic balance rather than static. The ECM is also influenced by cells, whether it is a parenchymal or non-parenchymal cell. None of the previous models explored the effects of mechanical factors on hepatocytes. To indicate it, the model that can change stiffness has been proposed. Fine-tuning of initial PEGdA/HA concentration and maintenance of stiffness close to normal liver parenchymal part would provide hepatocytes with prolonged functionally favorable microenvironments. Rapid degradation of HA-enriched zone in semi-IPNs by enzymatic activity and ECM formation by supporting cells leads to stiffness change. High-molecular-weight (1.5 MDa) HA formed defects in hydrogel networks during polymerization. After being degraded, HA-enriched zone provided a space for initial cell to spread and network forms. With the time going, the stiffness is always changing. The outcome when the elastic modulus of the 3D liver model converges close to that of the in vivo liver (≈2.3 to 5.9 kPa), both phenotypic and functional maturation of the 3D liver were realized [26]. Although these approaches were successful in the improvement of 3D hepatic cultures, biomechanical alterations and consequent functional changes in hepatic cells in the dynamic 3D microenvironment and tissue remodeling during long-term culture, are currently not well described.

**3.2.5 3D bio-printing:** 3D bio-printing has its fascinating advantages: convenience, precision, and is favored by scientists in the engineering and regenerative medicine. Just providing the biocompatible materials and cells, inputing the information about you need tissue into the system, then the machine will print it out, even for tiny things [66]. Vascular-like tubes, artificial skin, cartilage, and a wide range of tissue constructs also including stem cells are produced by various 3D printing technologies [67]. As for liver tissue, some researchers use DLP-based 3D bioprinting system to develop a 3D hydrogel-based triculture model that possesses the physiologically relevant cell combination and microarchitecture. The DLP-based 3D bioprinting system allows to embed hiPSC-HPCs and the supporting cells origin from both endothelial and mesenchymal in a 3D microscale hexagonal hydrogel construct [68].

**3.3 Vasculature formation**

**3.3.1 Blood vessel engineering:** This static culture condition lacks the shear stress associates with blood flow that is deemed to be a critical factor in determining cell fates [69, 70]. To address this problem, microfluidic was introduced into the design of liver cell culture model either by engineering "vascular-like" microfluidic structure pre-seeding or by promoting vasculature formation driven by the intrinsic properties of the seeded cells.

3D printing technology has been a powerful tool to engineer vascular-like microfluidic system, it allows us to fabricate device that mimics the liver anatomy like sinusoidal channels using artificial barrier layer [71]. On top of printing biomaterials as physical barrier, microfluidic direct writer (MFDW) enables the 3D printer to construct device using cell-laden hydrogel by using openings permitting media exchange [72]. The microfluidic fulfills the needs for physiological shear stress that is known to promote the primary liver cell's metabolic activity [70]. It has been theorized that the large increase in portal flow per unit of residual tissue mass may be among the earliest triggers of regeneration [52]. The results point to fluid mechanical stress and activation of regeneration-related genes in liver progenitor cells. Shear stress could upregulate regeneration-related immediate early genes in liver progenitors in 3D ECM-like microenvironments [73]. Each cell has its own most suitable shear force. The lower stress has no significant different from static culture and the higher shear stress in the flow had consequently detached the cells from the microcarriers, causing cell death and adverse cell count [69]. To fully utilize the benefit of having dynamic flow system that provide oxygen and nutrient to the hepatocytes, perfusion-incubator-liver-chip (PIC) has been developed to provide a long-term culture systems without manual medium change by assuring a tangential flow of the media over the spheroids culture. The essential attribute of PIC is that the design of the device assures a tangential flow over the cell culture to remove the metabolites and by-products from the proximity of the cells and refresh the cell environment [74].

**3.3.2 Cell-driven vasculature formation in organoid:** Besides the engineering approach mentioned above, in vitro-generated organ buds is a promising approach toward regenerating functional and vascularized organs. Takebe co-cultured human fetal liver cells (hFLCs) or hepatocytes derived from iPSCs with HUVECs and mesenchymal stem cells (hMSCs) and seeded these cells onto Matrigel. hMSCs served as the driving force to fold the co-culture from 2D sheet into 3D liver bud mimicking the fetal liver development. In the process of liver bud formation, HUVECs developed into premature vasculature networks in vitro. This result suggested that scaffold-free and self-condensation approaches are superior for the induction of vascularization [73, 75]. Although the presence of endothelial cells was dispensable for the generation of condensates, the post-transplant outcomes were clearly disappointing in the absence of HUVECs because no signs of functional vascularization were observed in vivo. As for hMSCs, they initiated condensation to form 3D structure while the lack of MSCs in the coculture led to a failure in condensate formation which was dependent upon soft substrate condition. In addition, the competition between cell-hydrogel and cell-cell interactions might be involved in the mechanism underlying the formation of cell condensates, especially in their initiation process [75]. By transplanting liver bud into mice, analyses of these implants may reveal the superiority of immature hepatic cells or progenitors for liver engineering rather than terminally differentiated mature hepatocytes. The authors also demonstrated the ability to scale up the production of liver buds to serve as the potential high throughput platform for drug discovery [76].

**3.4 Cell sourcing:** Since the stability and maintenance of the differentiated state of liver cells depends on both, the cell type used and the culture model (e.g. 2D cultures or complex 3D cultures), the choice of culture model in association with a specific hepatic cell source is critical for the success of individual hepatic in vitro studies.

Primary human hepatocytes (PHH) are regarded as the gold standard in vitro model to evaluate hepatic metabolism [77]. Because they naturally in the liver, reflect the complete functionality of the human organ and provide highly predictive results in toxicological in vitro research. Primary hepatocytes are preferable for cell therapy or pharmacology applications since they have not been altered by genetic mutations that lead to cancer and are the actual cell-type within the normal liver that catalyze exogenous medications. However, inter-individual differences and cell alterations due to the isolation procedure cause some variations in experimental results, which make the standardization of models difficult. In addition, the scarce availability and difficult logistics of primary human liver cells prevent a larger scale use of the cells. Primary hepatocytes are rarely available and can lose metabolic activity over the long term [38].

The immortalized cell line overcomes the shortcomings of primary hepatocytes and is widely used in drug metabolism and hepatotoxicity. The high expression of phase II genes makes the HepG2 cell line useful to study drug metabolisms [78]. Similar to HepG2, Huh7 cells express Phase I and Phase II xenobiotic drug metabolism genes, as well as hepatocyte-specific transcripts. However, alterations in the hepatocyte-specific functions limit the accuracy of results for humans because of transformation [77]. HepG2 cell line expresses very low levels of phase I drug-metabolizing enzymes such as cytochrome P450s [76] and the incomplete expression of surface junction proteins restricts hepatitis C virus entry into HepG2 that limit the usage in hepatisis exploration[79]. Huh7 human hepatoma-derived cell line allows the HCV entry and replication. At the same time, the disadvantage is that makes use of a non-differentiated cell line that does not recapitulate the cellular conditions encountered by HCV in vivo [80]. HepaRG, is a terminally differentiated hepatic cell derived from human hepatocellular carcinoma cell lines. Unlike HepG2 and Huh 7 lines, they possess stable phenotype and functional capacity of Phase I and II xenobiotic metabolizing enzymes and transporters over other hepatic cell lines. They could proliferate and then differentiate to hepatocytes and biliary cells under the right culture conditions [7]. Regardless of its superiority over other cancer cell lines, HepaRG is still suffering from subpar sensitivity in predicting drug metabolism and safety [81].

Although the claims on the identification and profiling of human liver cells have remained controversial, the conventional consensus on human fetal liver cells has always been set on two main populations, human hepatic stem cells (hHpSCs) and hepatoblasts. hHpSCs are AFP negative and known to be the precursors to hepatoblasts. HHpSCs have excellent proliferative capability, as shown by clonogenic expansion for >150 population doublings with phenotypic stability, and are pluripotent, with the ability to give rise directly to committed biliary progenitors and hepatoblasts, and thence to hepatocytic and biliary lineages, as well as to other endodermal cell types. Hepatoblasts are bipotent self-renewing cell type capable of differentiating into hepatocytes and cholangiocytes. The evidence that they yield mature liver tissue after transplantation supports the notion that they are potentially an excellent cell source for establishing long-term culture systems with advanced hepatic functions [34]. However, one big problem is that this cell source is hard to come by and the ethical concern is too much of a huddle to get approval from regulatory board.

Stem cells are the focus in the field of regeneration, such as induced pluripotent stem cells (iPSCs) from somatic cells by forced expression of the reprogramming factors Oct3/4 and Sox2 and human embryonic stem cells (hESCs) derived from the inner cell mass of blastocysts. ESCs and iPSCs can proliferate indefinitely without loss of potency and retain capacity for differentiation to various cells [8]. Nevertheless hESC raised ethical concerns about the use of hESC in research [81]. iPSCs have raised considerable excitement in the field of regeneration because this technology has the potential to fulfill the autologous transplantation that does not rely on the donor and life-long immunosuppression. Although various protocols in deriving hepatocytes using stem cell differentiation have been established to date, the generated hepatocyte-like cells (HLC) are characterized to be more of a fetal liver cells with immature phenotypes with reduced hepatic functionality than the fully functional PHH that are the gold standard for clinical application. In addition, due to the lack of standardized criteria and profiling assay, it is difficult to compare the success of different approaches and identify promising modifications that may enhance hepatocyte maturation [78]. All the existing protocol relies on the understanding of embryogenesis in guiding the supplementation of growth factors on the tissue culture dish [80].The advance of novel sequencing technology further fuel the development of differentiation protocol by identifying the right cell profiles and niches to produce.

**4. Future direction**

3D model is the trend now and will still be the main focus of liver research and development in foreseeable future. Bioartificial liver (BAL) system providing a life-saving straw for end-stage liver failure without liver transplant, is successful in supporting patient liver function, such as substance metabolism, detoxification, and albumin synthesis [82]. Injectable hydrogels have been paid more attention on cell therapy and tissue regeneration because of the applications in minimally invasive surgical procedures with ease of handling and complete filling of defect area [83]. Furthermore, organoids transplanted into the animal demonstrated rescue effect from acute liver failure and restored the critical liver function.

## 5. Summary

Phenotypical sustainability, authentic functionality are the main advantages of the 3D cell culture, which can provide great model for drug research and exploration of treatment. And more advanced 3D model come out with excellent characteristic, that brings positive significance to human development.

Reference

1. Lee, J., M.J. Cuddihy and N.A. Kotov, *Three-dimensional cell culture matrices: state of the art.* Tissue Eng Part B Rev, 2008. **14**(1): p. 61-86.

2. Kondo, Y., T. Iwao, K. Nakamura, T. Sasaki, S. Takahashi, N. Kamada, et al., *An Efficient Method for Differentiation of Human Induced Pluripotent Stem Cells into Hepatocyte-like Cells Retaining Drug Metabolizing Activity.* Drug Metabolism and Pharmacokinetics, 2014. **29**(3): p. 237-243.

3. Schwartz, R.E., H.E. Fleming, S.R. Khetani and S.N. Bhatia, *Pluripotent stem cell-derived hepatocyte-like cells.* Biotechnol Adv, 2014. **32**(2): p. 504-13.

4. Ebeling, A.H., *The Permanent Life of Connective Tissue Outside of the Organism.* J Exp Med,

1913. **17**(3): p. 273-85.

5. Periwal, V., J.R. Gaillard, L. Needleman and C. Doria, *Mathematical model of liver regeneration in human live donors.* J Cell Physiol, 2014. **229**(5): p. 599-606.

6. Strain, A.J. and J.M. Neuberger, *A bioartificial liver--state of the art.* Science, 2002. **295**(5557): p. 1005-9.

7. Andersson, T.B., K.P. Kanebratt and J.G. Kenna, *The HepaRG cell line: a unique in vitro tool for understanding drug metabolism and toxicology in human.* Expert Opinion on Drug Metabolism & Toxicology, 2012. **8**(7): p. 909-920.

8. Wang, Y.F., H.L. Yao, C.B. Cui, E. Wauthier, C. Barbier, M.J. Costello, et al., *Paracrine Signals from Mesenchymal Cell Populations Govern the Expansion and Differentiation of Human Hepatic Stem Cells to Adult Liver Fates.* Hepatology, 2010. **52**(4): p. 1443-1454.

9. Jiang, J., J.J. Briede, D.G.J. Jennen, A. Van Summeren, K. Saritas-Brauers, G. Schaart, et al., *Increased mitochondrial ROS formation by acetaminophen in human hepatic cells is associated with gene expression changes suggesting disruption of the mitochondrial electron transport chain.* Toxicology Letters, 2015. **234**(2): p. 139-150.

10. Mueller, D., L. Kramer, E. Hoffmann, S. Klein and F. Noor, *3D organotypic HepaRG cultures as in vitro model for acute and repeated dose toxicity studies.* Toxicol In Vitro, 2014. **28**(1): p. 104-12.

11. Kaneko, S., J. Furuse, M. Kudo, K. Ikeda, M. Honda, Y. Nakamoto, et al., *Guideline on the use of new anticancer drugs for the treatment of Hepatocellular Carcinoma 2010 update.* Hepatology Research, 2012. **42**(6): p. 523-542.

12. Prestwich, G.D., Y. Liu, B. Yu, X.Z. Shu and A. Scott, *3-D culture in synthetic extracellular*

*matrices: New tissue models for drug toxicology and cancer drug discovery.* Advances in Enzyme Regulation, Vol 47, 2007. **47**: p. 196-+.

13. DiMasi, J.A., R.W. Hansen and H.G. Grabowski, *The price of innovation: new estimates of drug development costs.* Journal of Health Economics, 2003. **22**(2): p. 151-185.

14. Ozenne, V., M. Bouattour, N. Goutte, M.P. Vullierme, M.P. Ripault, C. Castelnau, et al., *Prospective evaluation of the management of hepatocellular carcinoma in the elderly.* Dig Liver Dis, 2011. **43**(12): p. 1001-5.

15. Parkin, D.M., F. Bray, J. Ferlay and P. Pisani, *Global cancer statistics, 2002.* CA Cancer J Clin, 2005. **55**(2): p. 74-108.

16. Prestwich, G.D., *Evaluating drug efficacy and toxicology in three dimensions: Using synthetic extracellular matrices in drug discovery.* Accounts of Chemical Research, 2008. **41**(1): p. 139-148.

17. Skardal, A., L. Smith, S. Bharadwaj, A. Atala, S. Soker and Y.Y. Zhang, *Tissue specific synthetic ECM hydrogels for 3-D in vitro maintenance of hepatocyte function.* Biomaterials, 2012. **33**(18): p. 4565-4575.

18. Liehl, P., V. Zuzarte-Luis, J.N. Chan, T. Zillinger, F. Baptista, D. Carapau, et al., *Host-cell sensors for Plasmodium activate innate immunity against liver-stage infection.* Nature Medicine, 2014. **20**(1): p. 47-+.

19. Ringel, M., M.A. von Mach, R. Santos, P.J. Feilen, M. Brulport, M. Hermes, et al., *Hepatocytes cultured in alginate microspheres: an optimized technique to study enzyme induction.* Toxicology, 2005. **206**(1): p. 153-167.

20. Jitraruch, S., A. Dhawan, R.D. Hughes, C. Filippi, D. Soong, C. Philippeos, et al., *Alginate*

*Microencapsulated Hepatocytes Optimised for Transplantation in Acute Liver Failure.* Plos One, 2014. **9**(12).

21. LeCluyse, E.L., R.P. Witek, M.E. Andersen and M.J. Powers, *Organotypic liver culture models: meeting current challenges in toxicity testing.* Crit Rev Toxicol, 2012. **42**(6): p. 501-48.

22. Swift, B., N.D. Pfeifer and K.L. Brouwer, *Sandwich-cultured hepatocytes: an in vitro model to evaluate hepatobiliary transporter-based drug interactions and hepatotoxicity.* Drug Metab Rev, 2010. **42**(3): p. 446-71.

23. Turner, R., O. Lozoya, Y. Wang, V. Cardinale, E. Gaudio, G. Alpini, et al., *Human hepatic stem cell and maturational liver lineage biology.* Hepatology, 2011. **53**(3): p. 1035-45.

24. Suzuki, A., A. Iwama, H. Miyashita, H. Nakauchi and H. Taniguchi, *Role for growth factors and extracellular matrix in controlling differentiation of prospectively isolated hepatic stem cells.* Development, 2003. **130**(11): p. 2513-24.

25. Godoy, P., N.J. Hewitt, U. Albrecht, M.E. Andersen, N. Ansari, S. Bhattacharya, et al., *Recent advances in 2D and 3D in vitro systems using primary hepatocytes, alternative hepatocyte sources and non-parenchymal liver cells and their use in investigating mechanisms of hepatotoxicity, cell signaling and ADME.* Archives of Toxicology, 2013. **87**(8): p. 1315-1530.

26. Lee, H.J., M.J. Son, J. Ahn, S.J. Oh, M. Lee, A. Kim, et al., *Elasticity-based development of functionally enhanced multicellular 3D liver encapsulated in hybrid hydrogel.* Acta Biomaterialia, 2017. **64**: p. 67-79.

27. Li, C.Y., K.R. Stevens, R.E. Schwartz, B.S. Alejandro, J.H. Huang and S.N. Bhatia, *Micropatterned Cell-Cell Interactions Enable Functional Encapsulation of Primary*

*Hepatocytes in Hydrogel Microtissues.* Tissue Engineering Part A, 2014. **20**(15-16): p. 2200-2212.

28. Bhatia, S.N., G.H. Underhill, K.S. Zaret and I.J. Fox, *Cell and tissue engineering for liver disease.* Science Translational Medicine, 2014. **6**(245).

29. Walgren, J.L., M.D. Mitchell and D.C. Thompson, *Role of metabolism in drug-induced idiosyncratic hepatotoxicity.* Critical Reviews in Toxicology, 2005. **35**(4): p. 325-361.

30. Olson, H., G. Betton, D. Robinson, K. Thomas, A. Monro, G. Kolaja, et al., *Concordance of the toxicity of pharmaceuticals in humans and in animals.* Regulatory Toxicology and Pharmacology, 2000. **32**(1): p. 56-67.

31. Liu, Z.X., D. Han, B. Gunawan and N. Kaplowitz, *Neutrophil depletion protects against murine acetaminophen hepatotoxicity.* Hepatology, 2006. **43**(6): p. 1220-30.

32. Ochi, M., H. Ohdan, H. Mitsuta, T. Onoe, D. Tokita, H. Hara, et al., *Liver NK cells expressing TRAIL are toxic against self hepatocytes in mice.* Hepatology, 2004. **39**(5): p. 1321-31.

33. Brown, J.H., P. Das, M.D. DiVito, D. Ivancic, L.P. Tan and J.A. Wertheim, *Nanofibrous PLGA electrospun scaffolds modified with type I collagen influence hepatocyte function and support viability in vitro.* Acta Biomaterialia, 2018. **73**: p. 217-227.

34. Kim, B.S., I.K. Park, T. Hoshiba, H.L. Jiang, Y.J. Choi, T. Akaike, et al., *Design of artificial extracellular matrices for tissue engineering.* Progress in Polymer Science, 2011. **36**(2): p. 238-268.

35. Hadi, M., I.M. Westra, V. Starokozhko, S. Dragovic, M.T. Merema and G.M.M. Groothuis, *Human Precision-Cut Liver Slices as an ex Vivo Model to Study Idiosyncratic Drug-Induced Liver Injury.* Chemical Research in Toxicology, 2013. **26**(5): p. 710-720.

36. Xie, G.H., L. Wang, X.D. Wang, L. Wang and L.D. DeLeve, *Isolation of periportal, midlobular, and centrilobular rat liver sinusoidal endothelial cells enables study of zonated drug toxicity.* American Journal of Physiology-Gastrointestinal and Liver Physiology, 2010. **299**(5): p. G1204-G1210.

37. Ng, S.S., K. Saeb-Parsy, S.J.I. Blackford, J.M. Segal, M.P. Serra, M. Horcas-Lopez, et al., *Human iPS derived progenitors bioengineered into liver organoids using an inverted colloidal crystal poly (ethylene glycol) scaffold.* Biomaterials, 2018. **182**: p. 299-311.

38. Hussein, K.H., K.M. Park, J.H. Ghim, S.R. Yang and H.M. Woo, *Three dimensional culture of HepG2 liver cells on a rat decellularized liver matrix for pharmacological studies.* J Biomed Mater Res B Appl Biomater, 2016. **104**(2): p. 263-73.

39. Schuppan, D., M. Ruehl, R. Somasundaram and E.G. Hahn, *Matrix as a modulator of hepatic fibrogenesis.* Seminars in Liver Disease, 2001. **21**(3): p. 351-372.

40. Zhang, L.L., N. Theise, M. Chua and L.M. Reid, *The Stem Cell Niche of Human Livers: Symmetry Between Development and Regeneration.* Hepatology, 2008. **48**(5): p. 1598-1607.

41. Drury, J.L. and D.J. Mooney, *Hydrogels for tissue engineering: scaffold design variables and applications.* Biomaterials, 2003. **24**(24): p. 4337-4351.

42. Mooney, D., L. Hansen, J. Vacanti, R. Langer, S. Farmer and D. Ingber, *Switching from differentiation to growth in hepatocytes: control by extracellular matrix.* J Cell Physiol, 1992. **151**(3): p. 497-505.

43. Petrie, R.J., N. Gavara, R.S. Chadwick and K.M. Yamada, *Nonpolarized signaling reveals two distinct modes of 3D cell migration.* Journal of Cell Biology, 2012. **197**(3): p. 439-455.

44. Lee, W.C., C.H. Lim, Kenry, C. Su, K.P. Loh and C.T. Lim, *Cell-assembled graphene*

*biocomposite for enhanced chondrogenic differentiation.* Small, 2015. **11**(8): p. 963-9.

45. LeCouter, J., D.R. Moritz, B. Li, G.L. Phillips, X.H. Liang, H.P. Gerber, et al., *Angiogenesis-independent endothelial protection of liver: Role of VEGFR-1.* Science, 2003. **299**(5608): p. 890-893.

46. Friedman, S.L., *Hepatic stellate cells: Protean, multifunctional, and enigmatic cells of the liver.* Physiological Reviews, 2008. **88**(1): p. 125-172.

47. Henderson, T.M.A., K. Ladewig, D.N. Haylock, K.M. McLean and A.J. O'Connor, *Cryogels for biomedical applications.* Journal of Materials Chemistry B, 2013. **1**(21): p. 2682-2695.

48. Kolios, G., V. Valatas and E. Kouroumalis, *Role of Kupffer cells in the pathogenesis of liver disease.* World Journal of Gastroenterology, 2006. **12**(46): p. 7413-7420.

49. Nguyen, T.V., O. Ukairo, S.R. Khetani, M. McVay, C. Kanchagar, W. Seghezzi, et al., *Establishment of a hepatocyte-kupffer cell coculture model for assessment of proinflammatory cytokine effects on metabolizing enzymes and drug transporters.* Drug Metab Dispos, 2015. **43**(5): p. 774-85.

50. Wu, R., X. Cui, W. Dong, M. Zhou, H.H. Simms and P. Wang, *Suppression of hepatocyte CYP1A2 expression by Kupffer cells via AhR pathway: the central role of proinflammatory cytokines.* Int J Mol Med, 2006. **18**(2): p. 339-46.

51. Tacke, F., *Targeting hepatic macrophages to treat liver diseases.* J Hepatol, 2017. **66**(6): p. 1300-1312.

52. Zeng, T., C.L. Zhang, M. Xiao, R. Yang and K.Q. Xie, *Critical Roles of Kupffer Cells in the Pathogenesis of Alcoholic Liver Disease: From Basic Science to Clinical Trials.* Front Immunol, 2016. **7**: p. 538.

53. Hwang, C.M., S. Sant, M. Masaeli, N.N. Kachouie, B. Zamanian, S.H. Lee, et al., *Fabrication of three-dimensional porous cell-laden hydrogel for tissue engineering.* Biofabrication, 2010. **2**(3): p. 035003.

54. Carraro, A., W.M. Hsu, K.M. Kulig, W.S. Cheung, M.L. Miller, E.J. Weinberg, et al., *In vitro analysis of a hepatic device with intrinsic microvascular-based channels.* Biomedical Microdevices, 2008. **10**(6): p. 795-805.

55. Vernetti, L.A., N. Senutovitch, R. Boltz, R. DeBiasio, T.Y. Shun, A. Gough, et al., *A human liver microphysiology platform for investigating physiology, drug safety, and disease models.* Experimental Biology and Medicine, 2016. **241**(1): p. 101-114.

56. Kamalian, L., A.E. Chadwick, M. Bayliss, N.S. French, M. Monshouwer, J. Snoeys, et al., *The utility of HepG2 cells to identify direct mitochondrial dysfunction in the absence of cell death.* Toxicol In Vitro, 2015. **29**(4): p. 732-40.

57. Soars, M.G., D.F. McGinnity, K. Grime and R.J. Riley, *The pivotal role of hepatocytes in drug discovery.* Chemico-Biological Interactions, 2007. **168**(1): p. 2-15.

58. Chu, Q., Y. Zhao, X. Shi, W. Han, Y. Zhang, X. Zheng, et al., *In vivo-like 3-D model for sodium nitrite- and acrylamide-induced hepatotoxicity tests utilizing HepG2 cells entrapped in micro-hollow fibers.* Sci Rep, 2017. **7**(1): p. 14837.

59. Cerec, V., D. Glaise, D. Garnier, S. Morosan, B. Turlin, B. Drenou, et al., *Transdifferentiation of hepatocyte-like cells from the human hepatoma HepaRG cell line through bipotent progenitor.* Hepatology, 2007. **45**(4): p. 957-967.

60. De Kock, J., L. Ceelen, W. De Spiegelaere, C. Casteleyn, P. Claes, T. Vanhaecke, et al., *Simple and quick method for whole-liver decellularization: a novel in vitro three-dimensional* *bioengineering tool?* Arch Toxicol, 2011. **85**(6): p. 607-12.

61. Quante, M. and T.C. Wang, *Stem cells in gastroenterology and hepatology.* Nature Reviews Gastroenterology & Hepatology, 2009. **6**(12): p. 724-737.

62. Huang, Y., S. Onyeri, M. Siewe, A. Moshfeghian and S.V. Madihally, *In vitro characterization of chitosan-gelatin scaffolds for tissue engineering.* Biomaterials, 2005. **26**(36): p. 7616-27.

63. Huch, M. and L. Dolle, *The plastic cellular states of liver cells: Are EpCAM and Lgr5 fit for purpose?* Hepatology, 2016. **64**(2): p. 652-662.

64. German, C.L. and S.V. Madihally, *Type of endothelial cells affects HepaRG cell acetaminophen metabolism in both 2D and 3D porous scaffold cultures.* Journal of Applied Toxicology, 2019. **39**(3): p. 461-472.

65. Ng, S.S., A. Xiong, K. Nguyen, M. Masek, D.Y. No, M. Elazar, et al., *Long-term culture of human liver tissue with advanced hepatic functions.* JCI Insight, 2017. **2**(11).

66. Murphy, S.V. and A. Atala, *3D bioprinting of tissues and organs.* Nat Biotechnol, 2014. **32**(8): p. 773-85.

67. Lee, V., G. Singh, J.P. Trasatti, C. Bjornsson, X. Xu, T.N. Tran, et al., *Design and fabrication of human skin by three-dimensional bioprinting.* Tissue Eng Part C Methods, 2014. **20**(6): p. 473-84.

68. Marino, A., C. Filippeschi, V. Mattoli, B. Mazzolai and G. Ciofani, *Biomimicry at the nanoscale: current research and perspectives of two-photon polymerization.* Nanoscale, 2015. **7**(7): p. 2841-2850.

69. Ramaiahgari, S.C., M.W. den Braver, B. Herpers, V. Terpstra, J.N. Commandeur, B. van de

Water, et al., *A 3D in vitro model of differentiated HepG2 cell spheroids with improved liver-like properties for repeated dose high-throughput toxicity studies.* Arch Toxicol, 2014. **88**(5): p. 1083-95.

70. Esch, M.B., J.M. Prot, Y.I. Wang, P. Miller, J.R. Llamas-Vidales, B.A. Naughton, et al., *Multi-cellular 3D human primary liver cell culture elevates metabolic activity under fluidic flow.* Lab Chip, 2015. **15**(10): p. 2269-77.

71. Lee, P.J., P.J. Hung and L.P. Lee, *An artificial liver sinusoid with a microfluidic endothelial-like barrier for primary hepatocyte culture.* Biotechnol Bioeng, 2007. **97**(5): p. 1340-6.

72. Ghorbanian, S., M.A. Qasaimeh, M. Akbari, A. Tamayol and D. Juncker, *Microfluidic direct writer with integrated declogging mechanism for fabricating cell-laden hydrogel constructs.* Biomed Microdevices, 2014. **16**(3): p. 387-95.

73. Takebe, T., N. Koike, K. Sekine, R. Fujiwara, T. Amiya, Y.W. Zheng, et al., *Engineering of human hepatic tissue with functional vascular networks.* Organogenesis, 2014. **10**(2): p. 260-7.

74. Yu, F., R. Deng, W. Hao Tong, L. Huan, N. Chan Way, A. IslamBadhan, et al., *A perfusion incubator liver chip for 3D cell culture with application on chronic hepatotoxicity testing.* Sci Rep, 2017. **7**(1): p. 14528.

75. Takebe, T., M. Enomura, E. Yoshizawa, M. Kimura, H. Koike, Y. Ueno, et al., *Vascularized and Complex Organ Buds from Diverse Tissues via Mesenchymal Cell-Driven Condensation.* Cell Stem Cell, 2015. **16**(5): p. 556-65.

76. Takebe, T., K. Sekine, M. Kimura, E. Yoshizawa, S. Ayano, M. Koido, et al., *Massive and

*Reproducible Production of Liver Buds Entirely from Human Pluripotent Stem Cells.* Cell Rep, 2017. **21**(10): p. 2661-2670.

77. Zeilinger, K., N. Freyer, G. Damm, D. Seehofer and F. Knospel, *Cell sources for in vitro human liver cell culture models.* Experimental Biology and Medicine, 2016. **241**(15): p. 1684-1698.

78. Wilkening, S., F. Stahl and A. Bader, *Comparison of primary human hepatocytes and hepatoma cell line HEPG2 with regard to their biotransformation properties.* Drug Metabolism and Disposition, 2003. **31**(8): p. 1035-1042.

79. Schmelzer, E., L. Zhang, A. Bruce, E. Wauthier, J. Ludlow, H.L. Yao, et al., *Human hepatic stem cells from fetal and postnatal donors.* J Exp Med, 2007. **204**(8): p. 1973-87.

80. Si-Tayeb, K., F.K. Noto, M. Nagaoka, J. Li, M.A. Battle, C. Duris, et al., *Highly efficient generation of human hepatocyte-like cells from induced pluripotent stem cells.* Hepatology, 2010. **51**(1): p. 297-305.

81. Gerets, H.H.J., K. Tilmant, B. Gerin, H. Chanteux, B.O. Depelchin, S. Dhalluin, et al., *Characterization of primary human hepatocytes, HepG2 cells, and HepaRG cells at the mRNA level and CYP activity in response to inducers and their predictivity for the detection of human hepatotoxins.* Cell Biology and Toxicology, 2012. **28**(2): p. 69-87.

82. Streetz, K.L., *Bio-artificial liver devices - Tentative, but promising progress.* Journal of Hepatology, 2008. **48**(2): p. 189-191.

83. Tong, X.F., F.Q. Zhao, Y.Z. Ren, Y. Zhang, Y.L. Cui and Q.S. Wang, *Injectable hydrogels based on glycyrrhizin, alginate, and calcium for three-dimensional cell culture in liver tissue engineering.* Journal of Biomedical Materials Research Part A, 2018. **106**(12): p. 3292-3302.

84.  Wu, M., Z.H. Yang, Y.F. Liu, B. Liu and X.J. Zhao, *The 3-D Culture and In Vivo Growth of the Human Hepatocellular Carcinoma Cell Line HepG2 in a Self-Assembling Peptide Nanofiber Scaffold*. Journal of Nanomaterials, 2010.

85.  Malinen, M.M., L.K. Kanninen, A. Corlu, H.M. Isoniemi, Y.R. Lou, M.L. Yliperttula, et al., *Differentiation of liver progenitor cell line to functional organotypic cultures in 3D nanofibrillar cellulose and hyaluronan-gelatin hydrogels*. Biomaterials, 2014. **35**(19): p. 5110-5121

**Table1. The biomaterials used in the 3D models.**

**Fig. 1. Conventional 3D model**. (A) Cell encapsulation model: Cells and alginate solution mixture is dispersed into salt bath using syringe for rapid polymerization to form microbeads. (B) Sandwich model: A monolayer of hepatocytes is sandwiched by two layers of ECM protein such as collagen I or Matrigel. (C) Salt-leaching hydrogel: Sacrificial salt porogens are intitally mixed with the hydrogel for polymerization, then leached out to form macroporous hydrogel.

**Fig. 2. The structure of the liver hepatic lobules.** (A) The structure of the hepatic lobules and interlobules. (B) The location of the variety of the cells in liver.

**Fig. 3. Advanced 3D model.** (A) Decellularized liver: The liver organ is decellularized and reseeded using target cells for repopulation over the native matrix. (B) Inverted colloidal crystal (ICC): Hydrogel with highly uniform pore size and porosity mimicking the 3D hexagonal structure of liver

lobule. (C) Channel-liked scaffolds: cell suspensions are injected into the hollow fibers to form cylindroids.